\documentclass{mem}
\usepackage{natbib}\usepackage{txfonts}\usepackage{balance}
\usepackage{graphicx}
\usepackage[a4paper,breaklinks,dvipdfm]{hyperref}
\idline{83}{2}
\begin{document}
\def\teff{$T\rm_{eff }$}
\def\kms{$\mathrm {km s}^{-1}$}

\title{
The low states of CVs at the upper edge of the period gap
}

   \subtitle{}

\author{
P. \,Rodr\'\i guez-Gil\inst{1,2}, L. Schmidtobreick\inst{3}, K. S. Long\inst{4}\thanks{Visiting astronomer, Cerro Tololo Inter-American Observatory, National Optical Astronomy Observatory, which are operated by the Association of Universities for Research in Astronomy, under contract with the National Science Foundation.}, T. Shahbaz\inst{1,2},\\
B. T. G\"ansicke\inst{5} and M. A. P. Torres\inst{6} 
}

  \offprints{P. Rodr\'\i guez-Gil}

\institute{
Instituto de Astrof\'\i sica de Canarias, V\'\i a L\'actea s/n, La Laguna E-38205, Santa Cruz de Tenerife, Spain
\and
Departamento de Astrof\'\i sica, Universidad de La Laguna, La Laguna E-38204, Santa Cruz de Tenerife, Spain 
{\email{prguez@iac.es}
\and
European Southern Observatory, Casilla 19001, Santiago de Chile, Chile
\and
Space Telescope Science Institute, 3700 San Martin Drive, Baltimore, MD 21218, USA
\and
Department of Physics, University of Warwick, Coventry CV4 7AL, UK
\and
SRON, Netherlands Institute for Space Research, Sorbonnelaan 2, 3584 CA, Utrecht, the Netherlands}
}

\authorrunning{Rodr\'\i guez-Gil }

\titlerunning{The low states of CVs at the upper edge of the period gap}

\abstract{
We review our knowledge of the origin and phenomenology of low states (also known as VY Scl states) in cataclysmic variables, making special emphasis on the cataclysmic variable population found just above the period gap. This orbital period range between approximately 3 and 4 hours is the preferred land for the elusive SW Sextantis stars, which are believed to be about to enter the gap when their donor stars become fully convective. Despite their main role in our understanding of the whole picture of cataclysmic variable evolution, the study of their component stars is almost impossible due to the extreme veiling brightness of the accretion disc during the high state. Here we present the first steps toward the characterisation of the white dwarfs and the donor stars in these VY Scl systems in the low state, and the discovery of sporadic accretion events and satellite emission lines like those observed in the AM Her systems in the low state, which are likely related to magnetic activity of the fast rotating secondary stars.  
\keywords{binaries: close --
          stars: individual (HS\,0220+0603, BB Dor) --
          novae, cataclysmic variables}
}

\maketitle{}

\begin{figure*}[t]
\begin{center}
\includegraphics[width=6.3cm]{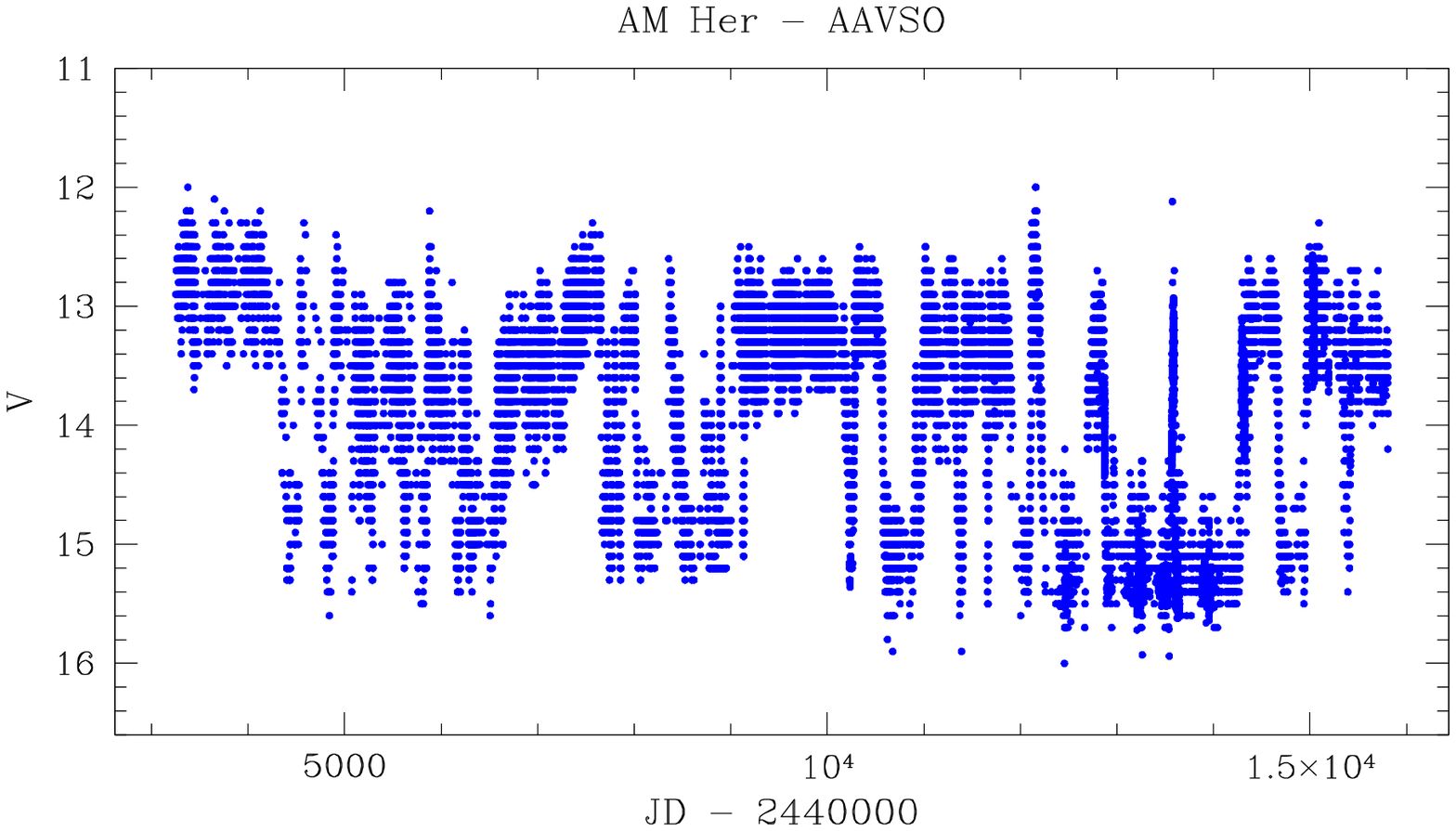}\includegraphics[width=6.3cm]{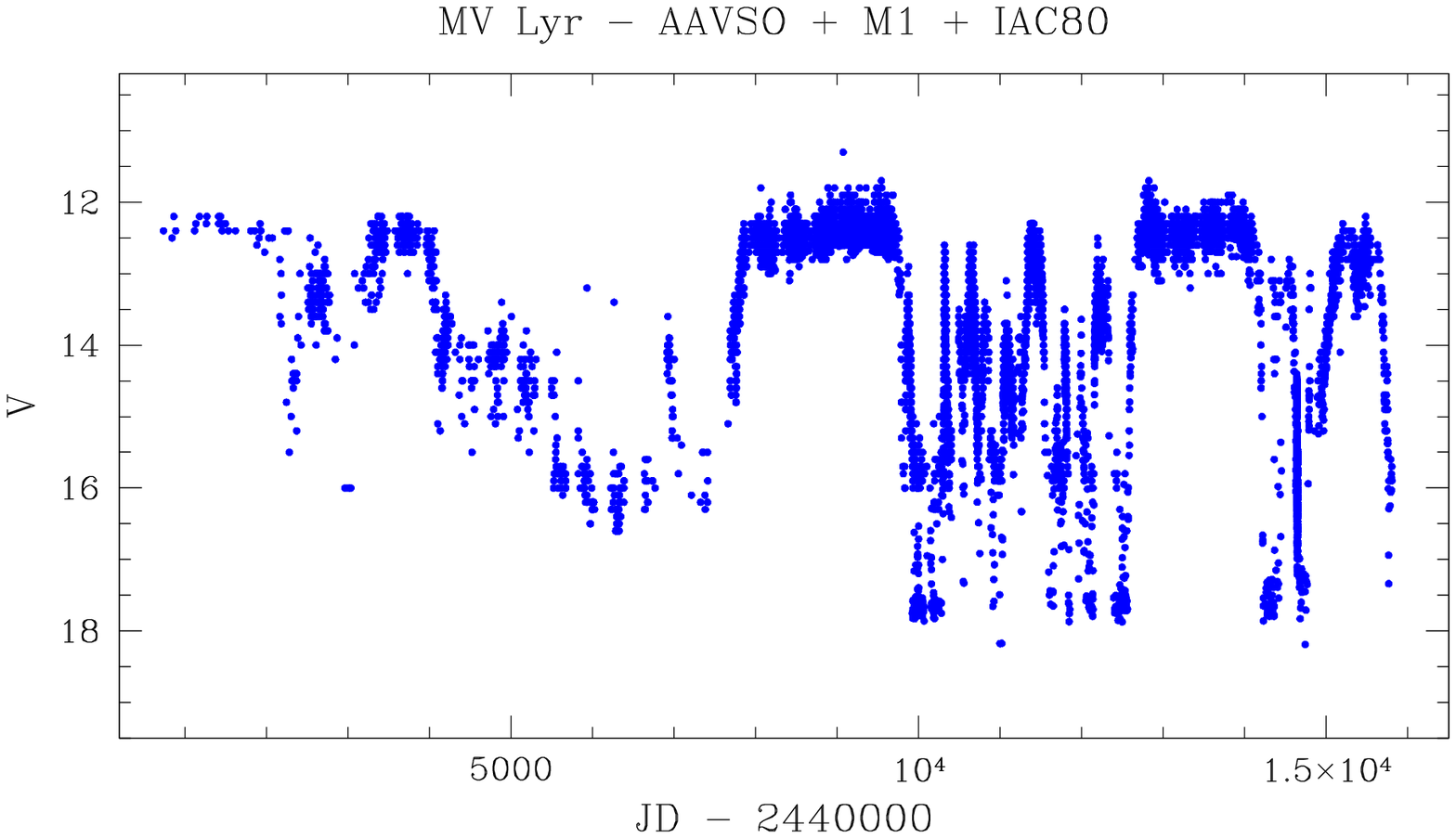}
\end{center}
\caption{
\footnotesize 34-year light curve of the polar CV prototype AM Her ({\it left}) and 37-year light curve of the nova-like MV Lyr ({\it right}). Both light curves are qualitatively very similar, indicating that the observed variability during the low states directly reflects changes in the mass transfer rate from the secondary star. Data from AAVSO, the Spanish M1 Group and the 0.82-m IAC80 telescope on Tenerife, Spain.}
\label{fig1}
\end{figure*}

\section{Introduction}

Cataclysmic variables (CVs) are occasionally caught fading toward states of greatly diminished brightness, or `low states' (also known as VY Scl states). During these unusual quiescent stages CVs shine $\sim$3-5 mag fainter than they do in the high state, and can stay at that level for days, months or even years before returning to the high state. Low states seem to occur independently of the magnetic field of the white dwarf (WD): they are observed in practically all AM Her stars (polars), some intermediate polars, and many weakly-magnetic CVs, including a large fraction of nova-likes and a number of dwarf novae and Z Cam stars. The causes and effects of low states have been previously reviewed by \cite{king+cannizzo98}, \cite{warner99}, and \cite{hessman00}.

The main driver of the low-state phenomenology came from the study of the discless AM Her stars. Without an accretion disc any large brightness drop must be caused by a significant decrease in the magnetic accretion luminosity, which is proportional to the mass transfer rate from the donor star. Therefore, the typical photometric changes seen during low states are directly related to variations in the rate at which the donor star provides matter.

The picture, however, is not as simple as a brightness drop followed by a relatively long quiescence phase and a return to the high state. As Fig.~\ref{fig1} shows CVs in the low state are always {\em eager} to recover, displaying quick brightness changes between the high and the low state, and even getting stuck at intermediate states for a while.

As mentioned, there is broad agreement that mass transfer from the donor star through the L1 point temporarily stops or greatly reduces during a low state, but the exact mechanism still remains unknown. \cite{livio+pringle94} proposed the accumulation of starspots close to the L1 point as a way of inhibiting Roche-lobe overflow. Only gas provided by the magnetic activity or stellar wind of the donor star would then be available for accretion \citep[e.g.][]{hessmanetal00}. In fact, entanglement of the magnetic fields of both stars has been proposed to explain the line emission patterns observed in AM Her during the low state \citep[][and references therein]{kafkaetal08}.

The observation of low states is therefore a unique opportunity to study the WDs and donor stars in CVs, especially the magnetic activity of the fast rotating donors, their solar-like activity cycles and the interplay with the strong WD magnetic field if present. On the other hand, the large mass accretion rate of nova-like CVs in the high state make their discs the dominating source of light, thus veiling the two stars and preventing dynamical measurements of their masses and other fundamental parameters such as the donor spectral type or the mean magnetic field of the WD from being done. It is only during deep low states that the WDs and the donor stars become visible for study.

\section{The population just above the period gap and CV evolution}

Of particular interest here are the low states of nova-like CVs which gather in the 3-4 h orbital period regime. They lie just at the upper boundary of the period gap, where donor stars are predicted to become fully convective and orbital angular momentum loss via magnetic wind braking is expected to greatly diminish or even cease \citep[e.g.][]{rappaportetal83}. Further, at least 50 per cent of all CVs in that period range belong to the SW Sex class \citep{rodriguez-giletal07}, characterized by having large mass accretion rates leading to very hot white dwarfs \citep{araujo-betancoretal05,townsley+gaensicke09}. As \citeauthor{townsley+gaensicke09} pointed out, no traditional wind braking angular momentum loss prescription can account for these large mass accretion rates, which makes the CVs populating the 3-4 h period region important targets to test the current CV evolution theory. In fact, work done by \cite{dantonaetal89} and \cite{dekool92} suggests a mass transfer turn-on at an orbital period of about 3.5 h as an explanation for the large accretion rates, and a pile-up of systems around the same period as observed, respectively.

\begin{figure*}[t!]
\begin{center}
\includegraphics[width=8.0cm]{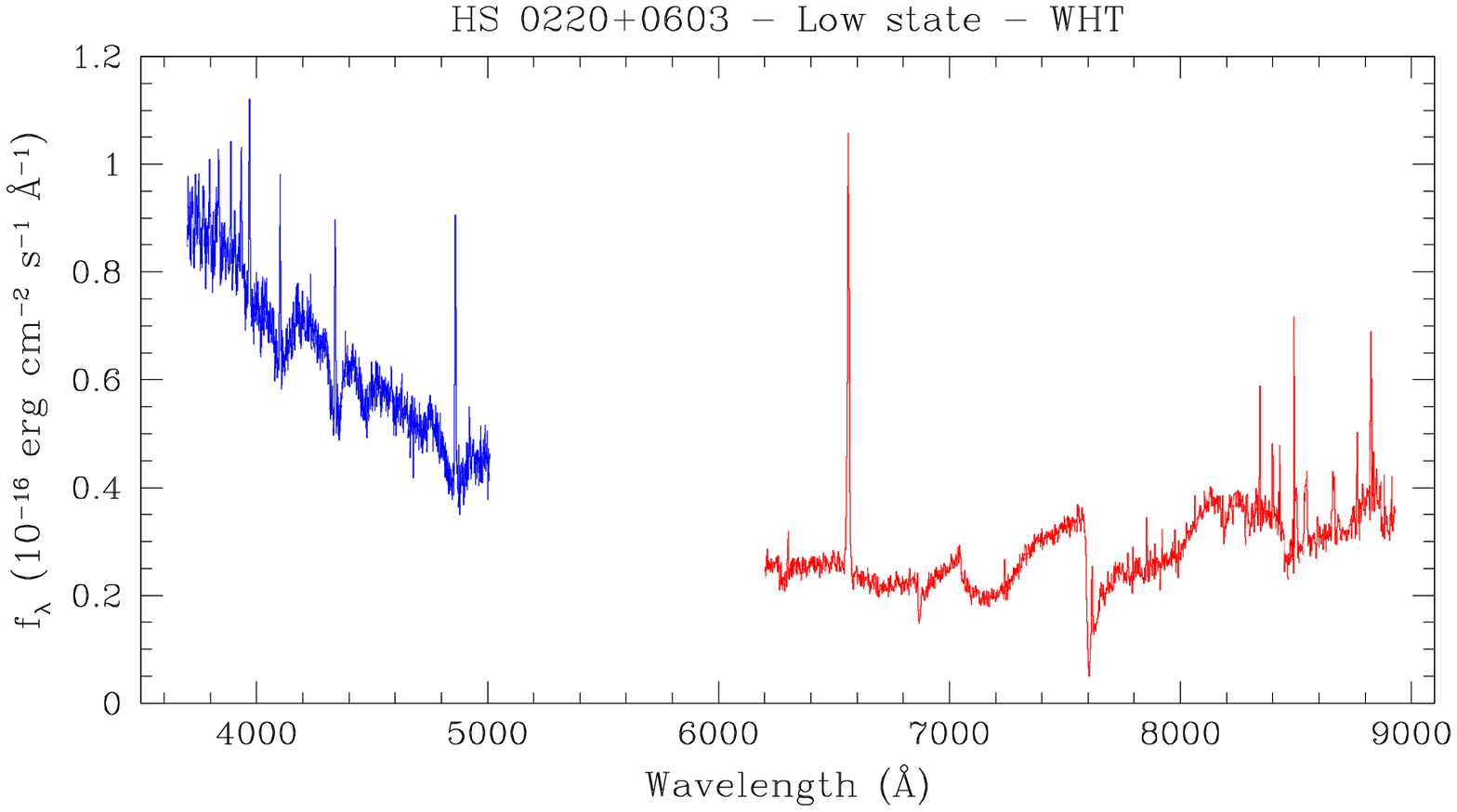}\\
\hspace{-0.3cm}
\includegraphics[width=7.6cm,angle=-90]{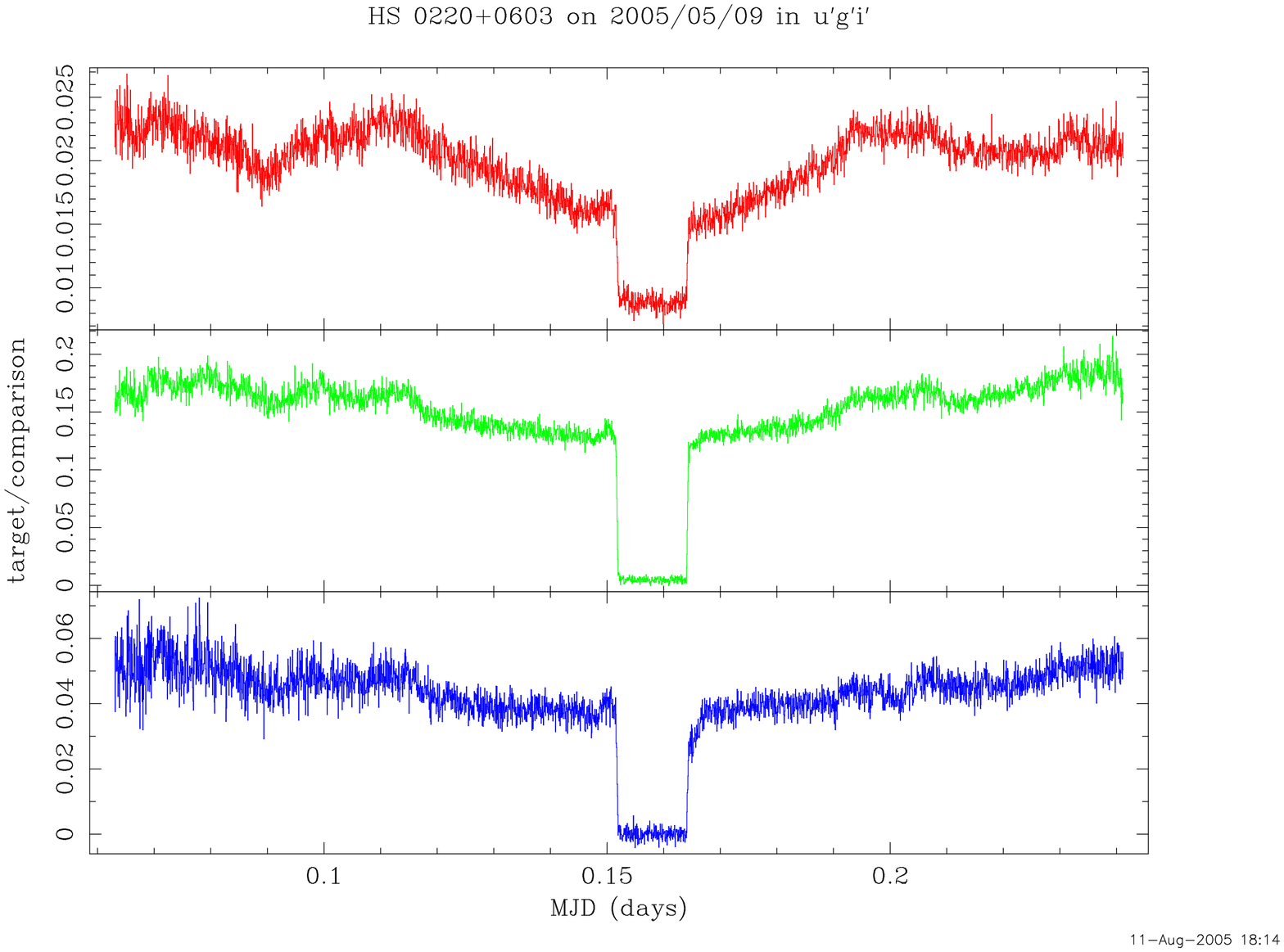}
\end{center}
\caption{
\footnotesize {\it Top}: Average spectrum of HS\,0220+0603 in the low state. {\it Bottom}: WHT/ULTRACAM light curves of the WD eclipse in the Sloan $u^\prime$ (blue), $g^\prime$ (green), and $i^\prime$ (red) bands.}
\label{fig2}
\end{figure*}

Another fundamental problem is the lack of observed dwarf nova outbursts in the VY Scl stars in the low state, which should take place once the accretion rate falls below a critical value. \cite{leachetal99} invoked irradiation of the inner accretion disc by a hot WD to suppress the outbursts for an unrealistic WD mass of 0.4 M$_\odot$. For a more plausible mass of 0.7 M$_\odot$ the outbursts would persist even at a WD temperature of 40\,000 K. \cite{hameury+lasota02} proposed the truncation of the inner disc by a magnetic WD as a means of quenching the outbursts. 

In this regard, we know that around half the nova-likes which have experienced a low state are SW Sex stars \citep{rodriguez-giletal07}. In addition, there is increasing evidence of significant magnetism in these systems from the detection of variable circular polarization \citep[e.g.][]{rodriguez-giletal01}, variable X-ray emission \citep{baskilletal05}, and emission-line flaring \citep[][and references therein]{rodriguez-giletal07}. The potential similarities between the behaviour in the low state of the strongly magnetic polar CVs and the SW Sex stars led us to start a long-term monitoring campaign to search for low states in the SW Sex systems and other VY Scl stars and perform time-resolved studies during these unique faint states. Here we present the first results of this project. 

\section{First dynamical mass measurement of an SW Sex star in the 3--4 h range}

\begin{figure*}[t!]
\begin{center}
\includegraphics[width=10cm,angle=-90]{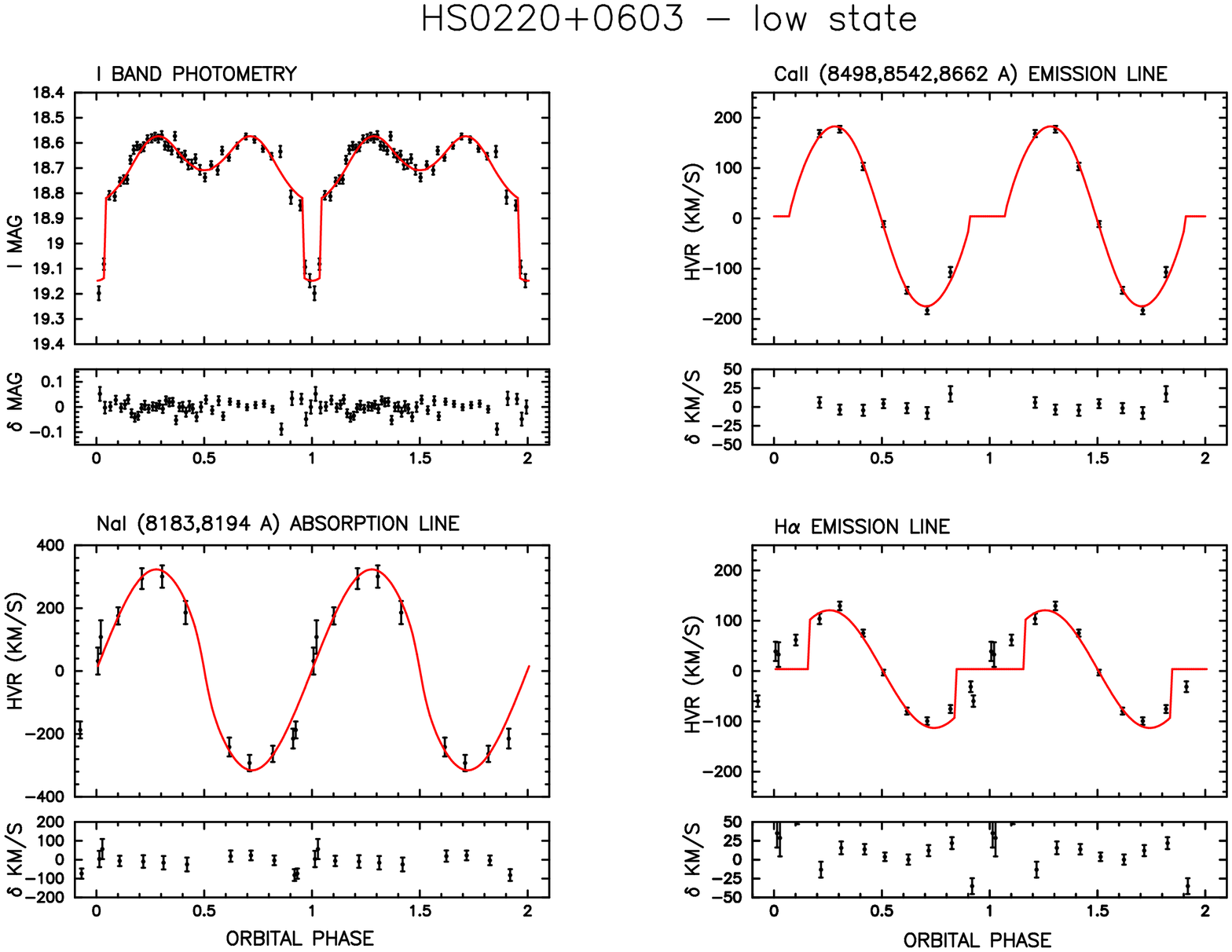}
\end{center}
\caption{
\footnotesize From {\it left} to {\it right} and from {\it top} to {\it bottom}: NOT $I$-band ellipsoidal+WD eclipse light curve;  Ca\,{\sc ii} emission triplet radial velocity curve; Na\,{\sc i} absorption doublet radial velocity curve; H$\alpha$ emission radial velocity curve. Red solid lines are the best fits with the exception of H$\alpha$ (model-predicted curve). The narrow plots under each panel are the corresponding residuals. An orbital cycle has been repeated.}
\label{fig3}
\end{figure*}

To answer the question whether the SW Sex stars and other VY Scl systems at the 3-4 h regime are about to enter the period gap or not (i.e. they may have born there), we need at least accurate masses on both sides of the gap. As no significant mass transfer is believed to occur within the gap, the stellar masses just above and below the gap should be quite similar. Results from precise light curve modelling of CVs below the gap yielded an average WD mass of $<M_1>\,=0.81 \pm 0.04$ M$_\odot$ and a mass of $M_2=0.18-0.22$ M$_\odot$ for the donors {\it just} below the gap \citep{savouryetal11}. \cite{kniggeetal11} reported a smaller value of $<M_1>\,=0.73\pm0.07$ M$_\odot$ below the gap, and $<M_1>\,=0.77\pm0.06$ M$_\odot$ above. However, the masses estimated for systems {\it just} above the gap are dubious since a variety of {\it a priori} assumptions are involved, the reason for this being the inability to detect the WD and the donor star in the high state (e.g. the eclipsing SW Sex stars DW UMa, \citealt{araujo-betancoretal03}, and UU Aqr, \citealt{diaz+steiner91}; \citealt{baptistaetal94}). Therefore, dynamical mass measurements during low states are needed.

We extensively observed the eclipsing SW Sex star HS\,0220+0603 \citep{rodriguez-giletal07} during its 2004--2005 low state (Rodr\'\i guez-Gil et al., in preparation). The data set comprises time-resolved spectroscopy and fast ULTRACAM photometry with the 4.2-m William Herschel Telescope (WHT), and $I$-band photometry with the 2.5-m Nordic Optical Telescope (NOT), both on La Palma. The WHT average spectrum, and the ULTRACAM light curves showing the eclipse of the WD are shown in Fig.~\ref{fig2}. Spectral modelling of this spectrum reveals a DAB WD with $T_\mathrm{eff} \simeq$ 30\,000 K and a M4--5 V donor star. The spectrum of HS\,0220+0603 also exhibits emission lines much narrower than in the high state, predominantly of the Balmer series, which come from the donor star as confirmed by their radial velocities. The near-IR sodium absorption doublet was used in combination with the motion of the Ca\,{\sc ii} triplet in emission to extract the radial velocity of the donor star. The velocities, together with the optical light curves were modelled according to the methods described by \cite{shahbazetal03}, which have successfully been used to model the light curves and radial velocity curves of neutron star and black hole X-ray binaries . Briefly, the model includes a Roche-lobe filling secondary star, the effects of heating of the secondary star by a source of high energy photons from the compact object, an accretion disc, and mutual eclipses of the disc and the secondary star.

In order to determine the binary system parameters we simultaneously fit the photometric light curve and the absorption-line (Na doublet) and emission-line (Ca\,{\sc ii} triplet) radial velocity curves model to represent the photometric and radial velocity variations. The best fits are presented in Fig.~\ref{fig3}. The H$\alpha$ emission radial velocity curve was excluded from the modelling as, unlike the Ca\,{\sc ii} triplet, H$\alpha$ emission is observed around zero phase. This indicates that not all the H$\alpha$ emission is driven by irradiation, chromospheric emission from the donor star may be significant. The system parameters are shown in Table~\ref{table_hs0220_fit}. The result of this first dynamical mass measurement in a SW Sex star yields a secondary star mass $\sim 40$ per cent larger than the mass predicted by the semi-empirical donor star sequence of \cite{kniggeetal11}. It's therefore clear that more dynamical mass measurements at the upper edge of the gap during low states are needed.  

\begin{table}[h!]
\caption{System parameters for HS\,0220+0603}
\label{table_hs0220_fit}
\begin{center}
\begin{tabular}{lc}
\hline
Parameter ~~~~~~~~~~~~~~~~~~~~~ & Value \\
\hline

\smallskip
$K_2$ (km s$^{-1}$)  & $284 ^{+14}_{-11}$  \\
\smallskip
$q$                             & $0.45 ^{+0.03}_{-0.04}$  \\
\smallskip
$i$ ($^{o}$)                 & 81  \\
\smallskip
$M_1/M_\odot$          & $0.77 ^{+0.13}_{-0.11}$  \\
\smallskip
$M_2/M_\odot$          & $0.34 ^{+0.09}_{-0.06}$  \\
\smallskip
$T_1$ (K)                  & 30\,000  \\
\smallskip
$T_2$ (K)                  & 3\,200  \\
\smallskip
$d$ (kpc)                     & $0.95 ^{+0.07}_{-0.10}$  \\
\hline
\end{tabular}
\end{center}
\end{table}

\begin{figure*}[t!]
\begin{center}
\includegraphics[width=9cm]{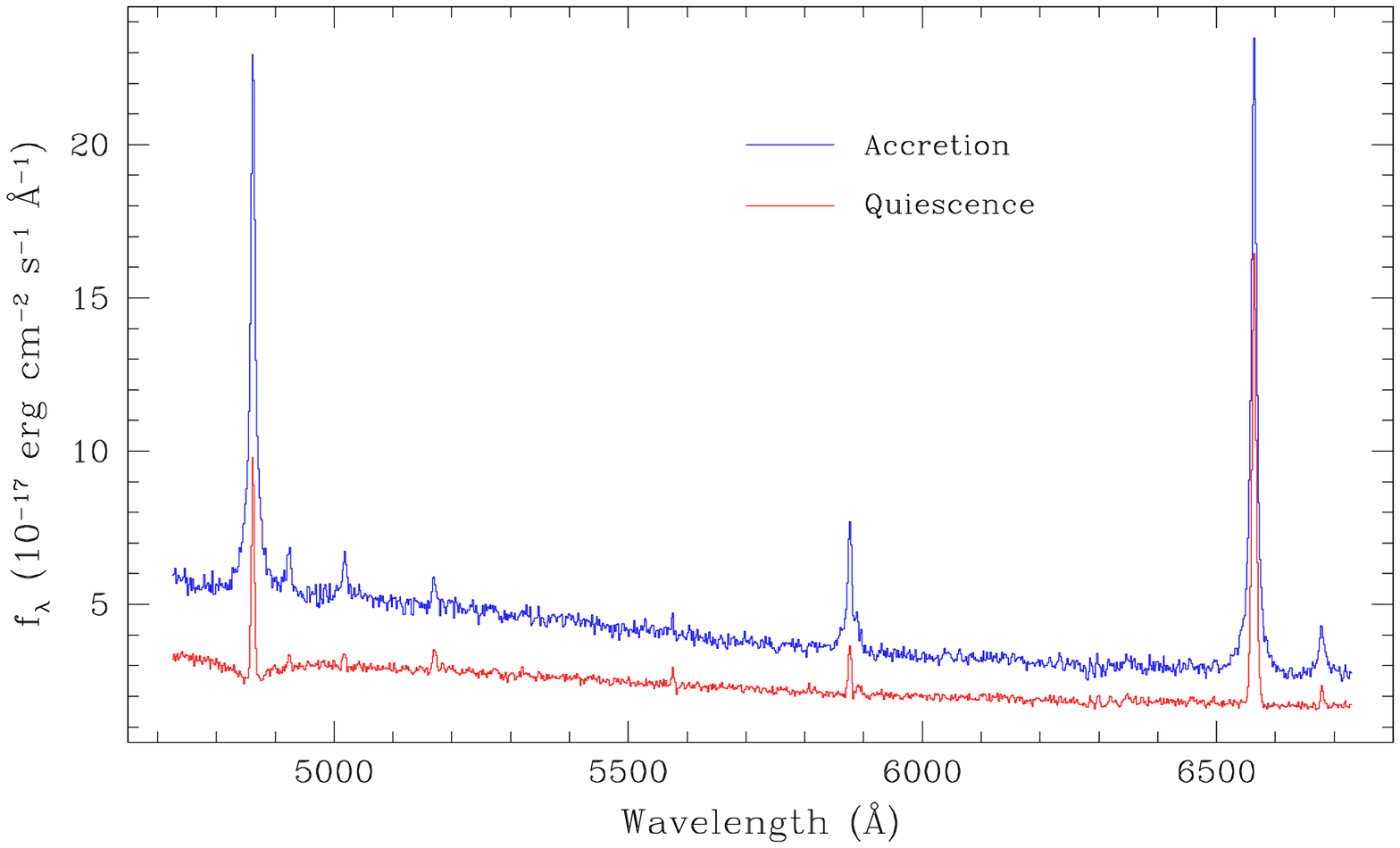}\\
\includegraphics[width=9.4cm]{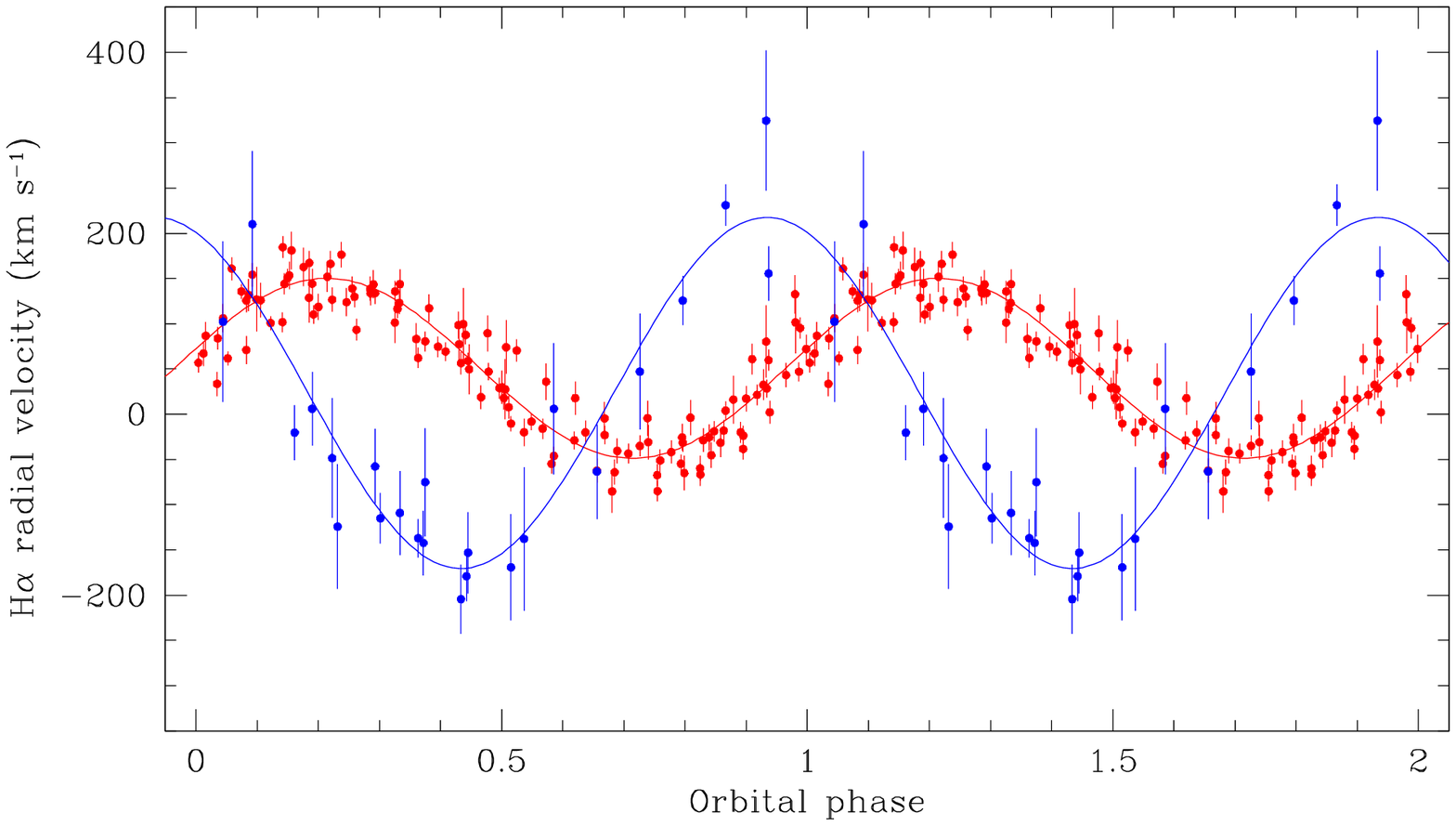}
\end{center}
\caption{
\footnotesize {\it Top}: Average of all quiescent spectra of BB Dor (in red) compared with the average of all the spectra taken during a total of 11 accretion events (in blue). {\it Bottom}: H$\alpha$ radial velocity curve of the line wings in the spectra obtained during the accretion events (blue) and H$\alpha$ radial velocity curve of the narrow profiles measured in the quiescent spectra.}
\label{fig4}
\end{figure*}

\section{Discovery of sporadic accretion events in BB Doradus}

Our long-term photometric monitoring of southern nova-like CVs with the 1.3-m SMARTS telescope found BB Doradus fading from $V \sim 14.3$ towards a deep low state at $V \sim 19.3$. Our NTT optical spectroscopy of BB Dor in this faint state revealed that the spectrum is a composite of a hot white dwarf ($T_\mathrm{eff} > 25\,000$ K) and a M3--4 donor star with narrow, mainly Balmer and He\,{\sc i} emission lines superposed. We associate this narrow profiles with line emission from the donor star. However, the spectrum of BB Dor in the low state shows significant variability. We detected episodic accretion events which veiled the absorption spectra of both stars and radically changed the line profiles within a timescale of tens of minutes. The narrow emission lines switched to broader lines with strong wings, and high-excitation lines such as He\,{\sc ii} $\lambda$4686 and the Bowen blend appeared. This shows that accretion is not complete quenched in the low state. In the top panel of Fig.~\ref{fig4} we show in blue the average of all the spectra taken during a total of 11 accretion events observed in 7 nights. The average of the remaining spectra (when no accretion was observed) is shown in red for comparison.

In the bottom panel of Fig.~\ref{fig4} we present the H$\alpha$ radial velocity curves of the strong line wings measured in the spectra obtained during the accretion episodes (in blue) and the narrow profiles measured in the quiescent spectra. The phase of the radial velocity curve of the H$\alpha$ emission-line wings indicates that the extra emission seen during the accretion events is not originated on the secondary star, but somewhere between the two stars. Remarkably, the wing velocities are delayed by 0.18 cycle with respect to the expected motion of the WD and reach their maximum in the blue at $\varphi \simeq 0.45$. This feature is a well-known characteristic of the emission S-wave observed both in the SW Sex stars \citep[see e.g.][]{rodriguez-giletal07} and the strongly magnetic polar CVs (i.e. AM Her stars) in the high state \citep[e.g. HU Aqr and V2301 Oph;][]{schwopeetal97,simicetal98}. This phase delay may suggest either connection of the material transferred from the secondary star to the magnetosphere of a magnetic WD in the absence of an accretion disc, or extra emission from the hot spot region fed by material hitting the outer rim of a cold, remnant accretion disc.

\section{Detection of H$\alpha$ satellite lines in BB Dor}

\begin{figure}
\includegraphics[width=7cm,angle=-90]{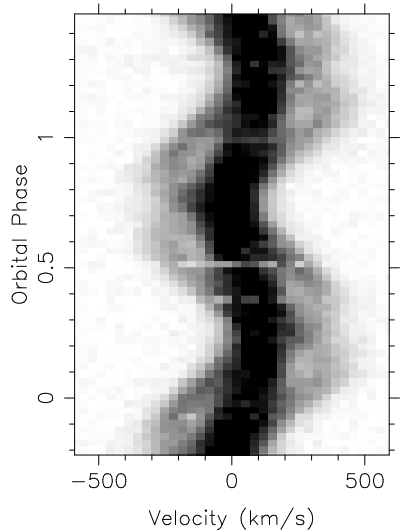}
\caption{
\footnotesize H$\alpha$ trailed spectra diagram showing the satellite lines and their orbital variation. Black corresponds to maximum emission.}
\label{fig5}
\end{figure}

We also observed BB Dor in the same low state with the VLT in Chile, just ten days after our NTT observations. The VLT time-resolved spectroscopy allowed a more detailed study of the H$\alpha$ emission line, and revealed a remarkable structure: the line profile consists of at least three emission components, a central emission line and two satellite lines. Their orbital variation is shown in the trailed spectra diagram in Fig.~\ref{fig5}, in which the two satellite lines can be clearly distinguished. They have the same radial velocity amplitude which is larger than that of the central emission component, and are symmetrically offset in phase by $\pm 0.15$.

These are the first ever observations of such satellite lines in an SW Sex star. However, similar satellite lines have been detected before in the low states of AM Her stars. The appearance of the satellite lines is strikingly similar in both types of CV \citep[][and references therein]{kafkaetal08}. Schmidtobreick et al. (in preparation) discuss the nature of the two satellite emissions in BB Dor, and suggest a likely origin in prominences created by the magnetically active secondary star.

\section{Conclusions}

Although we still don't exactly know what causes the low states in CVs, the observation with 8--10m class telescopes is providing new insights to the problem. Obtaining accurate stellar masses and radii in CVs thought to be close to enter the period gap is fundamental to test the current CV evolution theories. In addition, time-resolved, multifrequency observations during low states will help address open questions like the possible presence of a magnetic white dwarf or a remnant accretion disc in the CVs populating the 3--4 h orbital period region.

\begin{acknowledgements}
The use of Tom Marsh's \texttt{MOLLY} package is gratefully acknowledged. PRG thanks
the ESO/Santiago Visiting Scientist Program for the approval of a scientific visit
during which part of this work was completed. Partially funded by the Spanish MICINN under the Consolider-Ingenio 2010 Program grants CSD2006-00070: First Science with the GTC and CSD2009-00038: ASTROMOL.
Based in part on observations collected at the European Southern Observatory, La Silla and Paranal (programmes 082.D--0138 and 082.D--0154, respectively); on observations made with the William Herschel Telescope, which is operated on the island of La Palma by the Isaac Newton Group, and the Nordic Optical Telescope, operated on the island of La Palma jointly by Denmark, Finland, Iceland, Norway, and Sweden, both in the Spanish Observatorio del Roque de los Muchachos of the Instituto de Astrof\'\i sica de Canarias (IAC); and on observations with the IAC80 telescope, operated on the island of Tenerife by the IAC in the Spanish Observatorio del Teide.
\end{acknowledgements}

\bibliographystyle{aa}

\bigskip
\bigskip
\noindent {\bf DISCUSSION}

\bigskip
\noindent {\bf CHRISTIAN KNIGGE:} You showed the rapid development of broad H$\alpha$ wings in TT Ari. What was the timescale again?

\bigskip
\noindent {\bf PABLO RODR\'IGUEZ-GIL:} About twenty minutes.

\bigskip
\noindent {\bf CHRISTIAN KNIGGE:} So that means the development of the wings has nothing to do with the (re-)development of the disc.

\bigskip
\noindent {\bf PABLO RODR\'IGUEZ-GIL:} Right, it's much too fast.

\bigskip
\noindent {\bf GAGIK TOVMASSIAN:} You mentioned that most of the systems at the upper edge of the gap are either SW Sex or VY Scl type. I observed SS Aur which has a similar period but appears to be a dwarf nova most of the time, with regular outbursts, etc. However, during my observations it was in a very low state, unlike dwarf novae. So low I was able to see the WD and the red star, i.e. almost no disc. My comment and question is: maybe all objects at that period range have low states, but some of them experience it very infrequently. What do you think?

\bigskip
\noindent {\bf PABLO RODR\'IGUEZ-GIL:} I fully agree with you. The more we observe SW Sex/VY Scl stars, the larger the number of recorded low states. The recurrence times can be very long. The last low state of TT Ari took place almost 30 years after the previous one.

\end{document}